\documentclass{aastex631}

\usepackage{amsmath, amssymb}
\usepackage{graphicx}
\usepackage{multirow}
\usepackage{booktabs}

\newcommand{\comments}[1]{}

\shorttitle{Improving the Temporal Resolution of SOHO/MDI Magnetograms}
\shortauthors{Li et al.}

\begin{document}

\title{Improving the Temporal Resolution of SOHO/MDI Magnetograms 
of Solar Active Regions
Using a Deep Generative Model}

\correspondingauthor{Jason Wang, Haimin Wang}
\email{wangj@njit.edu, haimin.wang@njit.edu}

\author{Jialiang Li}

\affiliation{Institute for Space Weather Sciences, New Jersey Institute of Technology, University Heights, Newark, NJ 07102, USA}
\affiliation{Department of Computer Science, New Jersey Institute of Technology, University Heights, Newark, NJ 07102, USA}

\author{Vasyl Yurchyshyn}
\affiliation{Big Bear Solar Observatory, New Jersey Institute of Technology, 40386 North Shore Lane, Big Bear City, CA 92314, USA}

\author{Jason T. L. Wang}
\affiliation{Institute for Space Weather Sciences, New Jersey Institute of Technology, University Heights, Newark, NJ 07102, USA}
\affiliation{Department of Computer Science, New Jersey Institute of Technology, University Heights, Newark, NJ 07102, USA}

\author{Haimin Wang}
\affiliation{Institute for Space Weather Sciences, New Jersey Institute of Technology, University Heights, Newark, NJ 07102, USA}
\affiliation{Big Bear Solar Observatory, New Jersey Institute of Technology, 40386 North Shore Lane, Big Bear City, CA 92314, USA}
\affiliation{Center for Solar-Terrestrial Research, New Jersey Institute of Technology, University Heights, Newark, NJ 07102, USA}

\author{Yasser Abduallah}
\affiliation{Institute for Space Weather Sciences, New Jersey Institute of Technology, University Heights, Newark, NJ 07102, USA}
\affiliation{Department of Computer Science, New Jersey Institute of Technology, University Heights, Newark, NJ 07102, USA}

\author{Khalid A. Alobaid} 
\affiliation{Institute for Space Weather Sciences, New Jersey Institute of Technology, University Heights, Newark, NJ 07102, USA}
\affiliation{College of Applied Computer Sciences, King Saud University, Riyadh 11451, Saudi Arabia}

\author{Chunhui Xu}
\affiliation{Institute for Space Weather Sciences, New Jersey Institute of Technology, University Heights, Newark, NJ 07102, USA}
\affiliation{Department of Computer Science, New Jersey Institute of Technology, University Heights, Newark, NJ 07102, USA}

\author{Ruizhu Chen}
\affiliation{W.W. Hansen Experimental Physics Laboratory, Stanford University, Stanford, 
CA 94305, USA}

\author{Yan Xu}
\affiliation{Institute for Space Weather Sciences, New Jersey Institute of Technology, University Heights, Newark, NJ 07102, USA}
\affiliation{Big Bear Solar Observatory, New Jersey Institute of Technology, 40386 North Shore Lane, Big Bear City, CA 92314, USA}
\affiliation{Center for Solar-Terrestrial Research, New Jersey Institute of Technology, University Heights, Newark, NJ 07102, USA}

\begin{abstract}
We present a novel deep generative model,
named GenMDI,
to improve the temporal resolution of
line-of-sight (LOS) magnetograms
of solar active regions (ARs) collected by the Michelson Doppler Imager (MDI)
on board the Solar and Heliospheric Observatory (SOHO).
Unlike previous studies that focus primarily on spatial super-resolution of MDI magnetograms,
our approach can perform temporal super-resolution, which generates and inserts synthetic data between
observed MDI magnetograms, thus providing finer temporal structure
and enhanced details in the LOS data.
The GenMDI model employs a conditional diffusion process, 
which synthesizes images by considering both preceding and subsequent magnetograms, 
ensuring that the generated images are not only high-quality, 
but also temporally coherent with the surrounding data. 
Experimental results show that the GenMDI model
performs better than the traditional linear interpolation method,
especially in ARs with dynamic evolution in magnetic fields.
\end{abstract}

\section{Introduction} 
\label{sec:intro}

Deep-learning-based super-resolution of solar magnetograms
has drawn significant interest in recent years.
Previous studies focused on
data collected by
the Helioseismic and Magnetic Imager
\citep[HMI;][]{HMI} on board
the Solar Dynamics Observatory
\citep[SDO;][]{SDO} and
the Michelson Doppler Imager \citep[MDI;][]{MDI} 
on board the Solar and Heliospheric Observatory
\citep[SOHO;][]{SOHO}.
For example, \citet{2018A&A...614A...5D} developed
convolutional neural networks (CNNs) with
residual blocks to enhance HMI magnetograms.
\citet{2020ApJ...897L..32R} and
\citet{2021ApJ...923...76D} 
designed generative adversarial networks (GANs),
also to improve the spatial resolution of HMI data.
\citet{2022ApJS..263...25S} used
diffusion probabilistic models, combined with GANs
\citep{2024A&A...686A.272S},
to enhance the visible continuum images of HMI.
\citet{2024SoPh..299...36X}
developed an attention-aided CNN to enhance MDI
line-of-sight (LOS)
magnetograms.
Separately, \citet{2024ApJS..271....9D}
employed a multi-branch deep neural network to enhance the MDI LOS magnetograms.
\citet{2024ApJS..271...46M} used an encoder-decoder-based neural network
to improve MDI and GONG (Global Oscillation Network Group)
images to the HMI level.

All of the aforementioned studies aimed to improve
the spatial resolution of solar magnetograms.
None of the studies addressed the temporal super-resolution of solar magnetograms.
In this work, we made the first attempt
to improve the temporal resolution of SOHO/MDI LOS magnetograms of solar active regions (ARs).
The nominal MDI cadence is 96 minutes, far from sufficient to study dynamic evolution of solar magnetic fields in ARs.
Higher temporal resolution of the magnetograms helps scientists better
understand the dynamics and evolution of solar ARs
in the MDI era.

We adopt a deep generative model, specifically a diffusion model,
named GenMDI,
for the temporal super-resolution of MDI magnetogram images.
Diffusion models
can synthesize images, generate speech, and process video, among others.
Normally, 
these models work by inverting the process of natural diffusion, 
where they start with a distribution of random noise and progressively transform it
into a structured data distribution resembling the training data.
This transformation occurs in multiple steps,
which incrementally denoise the noisy sample until it reaches the desired complexity and detail. 

In contrast to the normal diffusion models mentioned above
\citep{2022ApJS..263...25S,2024A&A...686A.272S},
which generate synthetic images by
denoising random noise distributions without incorporating any specific
guidance, our GenMDI model generates a synthetic image considering the previous image and the next image surrounding the generated image. This image generation process
with guidance or condition is known as the
conditional diffusion process,
which is often used in the generation of video frames
\citep{DBLP:conf/nips/VoletiJP22}.
By conditioning the reverse diffusion process on the previous and subsequent images,
GenMDI ensures that the generated image maintains continuity and reflects the dynamics of the surrounding images. 
This approach not only preserves the natural flow and consistency of MDI
time-series magnetograms
but also enhances our model's ability to accurately generate synthetic images.
To our knowledge, this is the first time a conditional diffusion model
has been used to improve the temporal resolution of MDI magnetograms.

The remainder of this paper is organized as follows.
Section \ref{sec:data} describes the data used in this study.
Section \ref{sec:method}
details our conditional diffusion model (GenMDI),
explaining how the model works.
Section \ref{sec:results} reports the experimental results.
Section \ref{sec:conclusion} concludes the article.

\section{Data}
\label{sec:data}

We adopted the data product, named Space-Weather MDI
Active Region Patches \citep[SMARPs;][]{2021ApJS..256...26B}, 
derived from maps of the
solar surface magnetic field taken by SOHO/MDI.
This data product is available from
the Joint Science Operations Center
(JSOC) at \url{http://jsoc.stanford.edu/}.
We downloaded the MDI LOS magnetograms from the
{\sf mdi.smarp\_cea\_96m} data series at the JSOC site.
These magnetograms are available from 1996 to 2010 at a cadence of 96 min.

MDI CEA magnetograms were selected in 237 ARs and
split into a training set (202 ARs) and a test set (35 ARs). 
In total, there are 4222 magnetograms in the training set
and 483 magnetograms in the test set.
The ARs in the training set and those in the test set are disjoint, 
and hence
our GenMDI model is tested on the data that
the model has never seen during the training process.

GenMDI is a conditional diffusion model that works
by considering the previous image and the next image of
a generated image.
Thus, in constructing training data, 
we use a sliding-window approach to create triplets of images
$(\mathbf{w}, \mathbf{x}, \mathbf{y})$ in which $\mathbf{x}$ is a ground truth image,
$\mathbf{w}$ is its previous image and $\mathbf{y}$ is its next image
in a sequence of training MDI magnetogram images.
When there is a missing image, we ignore the missing image and
slide to its right position to take the next available image to form a triplet.
In this way, we create 3634 triplets of magnetogram images in the training set.
Using the same method, we create 392 triplets of magnetogram images in the test set.
Table \ref{table:data} summarizes the numbers of ARs, magnetograms, and triplets of images in the training set and the test set, respectively.

\begin{table}
\centering
\hspace*{-1.9cm}
\begin{tabular}{|c|c|c|c|}
\hline
 Data Set            & No. of ARs  & No. of Magnetograms 
 & No. of Triplets \\
\hline
Training Set & 202 & 4222        & 3634    \\ \hline
Test Set  & 35  & 483        & 392    \\ \hline
\end{tabular}
\caption{Summary of SMARP Data Sets Used in Our Study}
\label{table:data}
\end{table}

\section{Methodology}
\label{sec:method}

\begin{figure*}
    \centerline{
     \includegraphics[width=0.805\columnwidth]{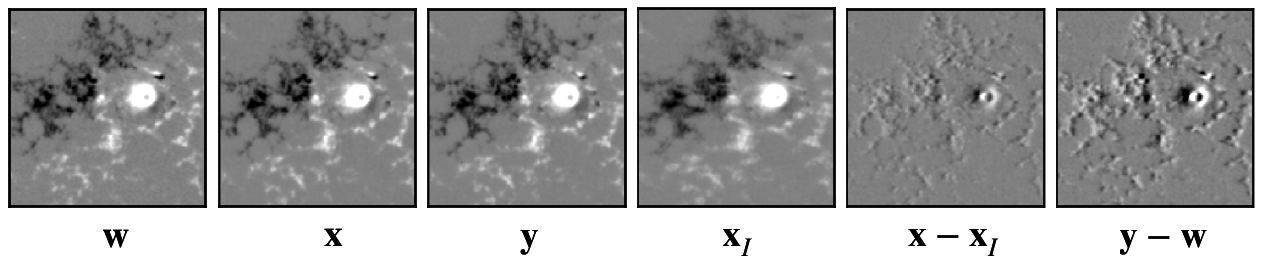}
    }
    \centerline{
      \hspace{0\textwidth}  \color{black}{\normalsize ({a})}
      }
    \centerline{
     \includegraphics[width=0.80\columnwidth]{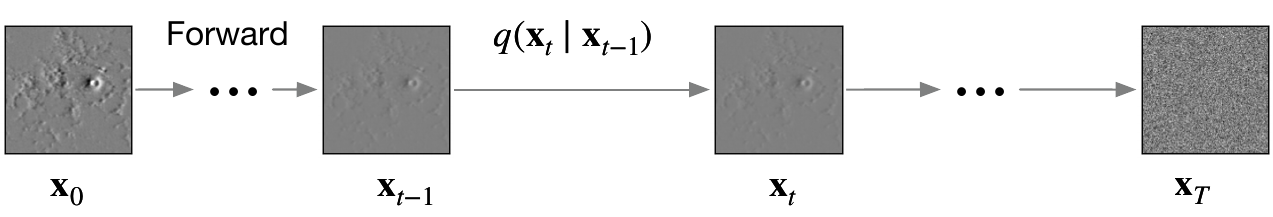}
    }
    \centerline{
      \hspace{0\textwidth}  \color{black}{\normalsize ({b})}
      }
    \centerline{
     \includegraphics[width=0.80\columnwidth]{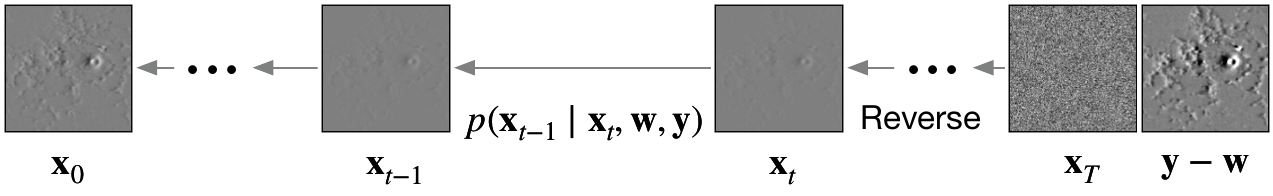}
    }
    \centerline{
      \hspace{0\textwidth}  \color{black}{\normalsize ({c})}
      }
    \vspace{0.02\textwidth}
    \vspace{0.01\textwidth}
    \caption{Illustration of the forward and reverse diffusion processes.
   (a) For a triplet of images $(\mathbf{w}, \mathbf{x}, \mathbf{y})$, we calculate the difference images 
    $(\mathbf{x} - \mathbf{x}_{I})$ and
    $(\mathbf{y}-\mathbf{w})$, where $\mathbf{x}_{I}$ is obtained
by applying linear interpolation to $\mathbf{w}$ and $\mathbf{y}$.
    (b) In the forward diffusion process, we gradually add noise to 
    the original difference image $\mathbf{x}_0$ = 
    $(\mathbf{x} - \mathbf{x}_{I})$ to transform it to the Gaussian noise 
    $\mathbf{\mathbf{x}}_T$. 
    (c) In the reverse diffusion process, 
    we start from $\mathbf{x}_T$ along with the conditional image $(\mathbf{y}-\mathbf{w})$ and attempt to denoise
    $\mathbf{x}_T$ into the original difference image
$\mathbf{x}_0$.}
\label{fig:normaldiff}
\end{figure*}
A diffusion model consists of a forward diffusion process
and a reverse diffusion process.
In the forward diffusion process, we progressively
corrupt the input data, transitioning from the original distribution to a
Gaussian distribution over a predetermined number of steps $T$. 
(In the study presented here, $T$ is set to 999.)
We build GenMDI through linear interpolation,
which is a widely used data interpolation method, as follows.
Consider a triplet of images $(\mathbf{w}, \mathbf{x}, \mathbf{y})$ in the training set
(test set, respectively).
Let $\mathbf{x}_{I}$ represent the image obtained
by applying linear interpolation to $\mathbf{w}$ and $\mathbf{y}$.
The difference image between $\mathbf{x}$ and $\mathbf{x}_{I}$ 
is denoted as $(\mathbf{x} - \mathbf{x}_{I})$.
We model the corruption of
$\mathbf{x}_0$ = 
$(\mathbf{x} - \mathbf{x}_{I})$ at time step 0 (i.e., $t=0$)
by gradually adding noise over a series of steps, from $t=1$ to $t=T$. 
This process is defined by the following transition
kernel \citep{DBLP:conf/nips/HoJA20}:
\begin{equation}
q(\mathbf{x}_t \mid \mathbf{x}_{t-1}) = 
\mathcal{N}(\mathbf{x}_t; \sqrt{1-\beta_t} \mathbf{x}_{t-1}, \beta_t \mathbf{I}),
\end{equation}
where
$\mathbf{x}_t$ is the difference image at step $t$,
$\mathbf{x}_{t-1}$ is the difference image 
at step $t-1$,
$\mathcal{N}$ denotes the normal/Gaussian
distribution,
$\beta_t$ is the variance schedule of the noise
added at step $t$, 
and
$\mathbf{I}$ is the 
identity
matrix. 
The cumulative effect of the noise across all steps up to $t$ 
can be represented by the accumulated variable
$\bar{\alpha}_t$, calculated as follows:
\begin{equation}
\bar{\alpha}_t = \prod_{s=1}^t(1-\beta_s), 
\end{equation}
which allows us to express the corrupted difference image at step $t$
directly in terms of the original difference image $\mathbf{x}_0$, i.e.,
\begin{equation}
q(\mathbf{x}_t \mid \mathbf{x}_0) = \mathcal{N}(\mathbf{x}_t; 
\sqrt{\bar{\alpha}_t} \mathbf{x}_0, (1-\bar{\alpha}_t) \mathbf{I}).
\label{forward}
\end{equation}
Equation (\ref{forward}) can be rewritten as
\citep{DBLP:conf/nips/HoJA20}:
\begin{equation}
\mathbf{x}_t = \sqrt{\bar{\alpha}_t} \mathbf{x}_0 + 
\sqrt{1 - \bar{\alpha}_t} \boldsymbol{\epsilon}_t,
\end{equation}
where $\boldsymbol{\epsilon}_t \sim \mathcal{N}(\mathbf{0}, \mathbf{I})$
is the noise added in step $t$.
The reverse diffusion process starts from
 $\mathbf{x}_T$, 
 along with the conditional image
 $(\mathbf{y} - \mathbf{w})$,
and attempts to denoise $\mathbf{x}_T$ into  
the original difference image
$\mathbf{x}_0$.
Figure \ref{fig:normaldiff} illustrates the forward and reverse diffusion processes.
The goal of training our GenMDI model is to learn the reverse diffusion process, 
i.e., training
$p(\mathbf{x}_{t-1} \mid \mathbf{x}_t, \mathbf{w}, \mathbf{y})$.
We adopt an enhancement of U-Net \citep{falk2019u},
named U-Net$^{+},$ 
for our model training and testing.

\begin{figure}
\centering
\includegraphics[width=0.60\linewidth]{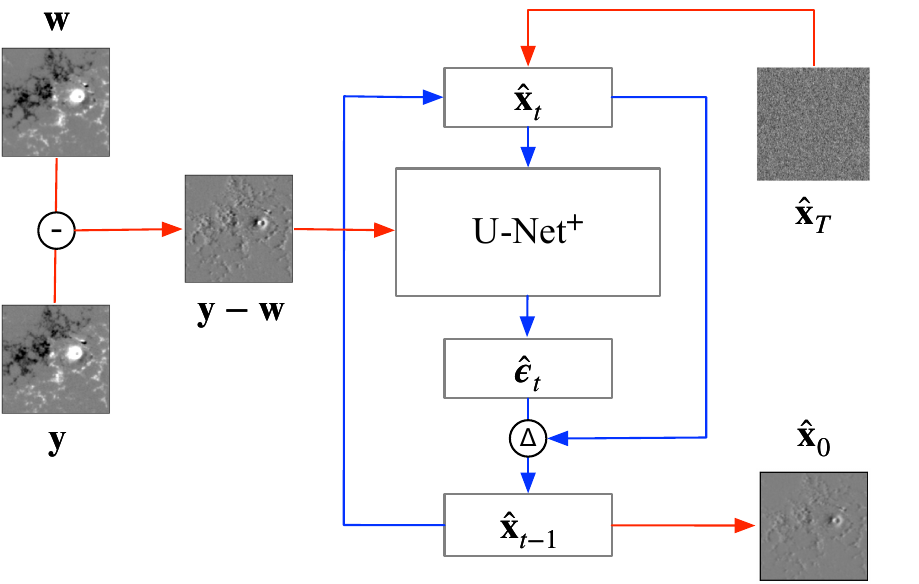}
\caption{
The inference process of GenMDI.
The trained U-Net$^{+}$ model in GenMDI takes as input 
$\hat{\mathbf{x}}_T$, which is a
randomly generated Gaussian noise image,  
along with the conditional image
 $(\mathbf{y} - \mathbf{w})$, and
predicts as output a synthetic difference image
$\hat{\mathbf{x}}_0$, 
which approximates
the original difference image $\mathbf{x}_0$ = 
    $(\mathbf{x} - \mathbf{x}_{I})$ 
    in Figure \ref{fig:normaldiff}.
The conditional image $(\mathbf{y} - \mathbf{w})$
is used to guide the model inference process.
Note that red lines represent the data flow of the input/output images. Blue lines represent the control flow of the iterative procedure used by the GenMDI model.
}
\label{fig:diff_architecture}
\end{figure} 

During training,
we feed all triplets of training images in the training set
to the model.
For a triplet of training images
$(\mathbf{w}, \mathbf{x}, \mathbf{y})$, 
we randomly choose an image
$\mathbf{x}_t$, $1 \leq t \leq T$,
obtained from the forward diffusion process.
We feed $\mathbf{x}_t$ as input to U-Net$^{+},$
which predicts as output noise $\hat{\boldsymbol{\epsilon}}_t$
conditioned on the difference image
$(\mathbf{y} - \mathbf{w})$.
The predicted noise $\hat{\boldsymbol{\epsilon}}_t$
is compared with the $\boldsymbol{\epsilon}_t$ generated in the forward diffusion process.
We adopt the loss function ${\cal L}$, defined as follows:
\begin{equation}
{\cal L} = \mathbb{E}_{t, [\mathbf{w}, \mathbf{x}, \mathbf{y}] \sim p_{\text{data}}, 
\boldsymbol{\epsilon}_t \sim \mathcal{N}(\mathbf{0}, \mathbf{I})}
\left[\left\|\boldsymbol{\epsilon}_t - 
\hat{\boldsymbol{\epsilon}}_t\right\|^2\right],
\label{loss_function}
\end{equation}
where $\mathbb{E}$ is the expectation value, and
$\mathbf{w}$, $\mathbf{x}$, $\mathbf{y}$ are from the data distribution 
$p_{\text{data}}$.
The loss function is optimized so that
$\hat{\boldsymbol{\epsilon}}_t$ approximates
$\boldsymbol{\epsilon}_t$.
The model training
is completed after sufficient
images $\mathbf{x}_t$ are chosen and the corresponding
noise values $\hat{\boldsymbol{\epsilon}}_t$
are optimized.

During inference/testing, 
we consider each triplet of test images
$(\mathbf{w}, \mathbf{x}, \mathbf{y})$, one at a time, in the test set.
Each time, 
we feed
$(\mathbf{w}, \hat{\mathbf{x}}_{T}, \mathbf{y})$ as input to the trained GenMDI model,
where
$\hat{\mathbf{x}}_T$ is a Gaussian noise image generated randomly
(see Figure \ref{fig:diff_architecture}).
As in the training process, 
we use the difference image
$(\mathbf{y} - \mathbf{w})$
as the conditional image, which is used
to guide the model inference process.
Initially, step $t$ = $T$, and we feed
$\hat{\mathbf{x}}_t$ = $\hat{\mathbf{x}}_T$ 
to the trained U-Net$^{+}$.
The trained U-Net$^{+}$ predicts as output noise $\hat{\boldsymbol{\epsilon}}_t$.
Next, we obtain $\hat{\mathbf{x}}_{t-1}$
by denoising $\hat{\mathbf{x}}_t$ using $\hat{\boldsymbol{\epsilon}}_t$,
where the denoising operator is represented by
$\bigtriangleup$.
The obtained $\hat{\mathbf{x}}_{t-1}$  
becomes
$\hat{\mathbf{x}}_{t}$ in the next iteration.
Through the iterative procedure,
the GenMDI model transforms
$\hat{\mathbf{x}}_T$
into a synthetic difference image
$\hat{\mathbf{x}}_{0}$.
Finally, the model generates as output the synthetic image
$\hat{\mathbf{x}}$
 = $\mathbf{x}_{I}$ + $\hat{\mathbf{x}}_0$. 
 In other words, GenMDI learns and predicts
the synthetic difference image
$\hat{\mathbf{x}}_0$
and then produces as output the synthetic image $\hat{\mathbf{x}}$
by adding $\hat{\mathbf{x}}_0$ to $\mathbf{x}_{I}$
where $\hat{\mathbf{x}}$ approximates the true image $\mathbf{x}$.

The U-Net$^{+}$ in Figure \ref{fig:diff_architecture}
has an encoder, a bottleneck, and
a decoder, as in U-Net \citep{falk2019u}.
The encoder consists of six layers
(E1, E2, E3, E4, E5, E6), and
the decoder also has six layers
(D1, D2, D3, D4, D5, D6).
Each encoder layer contains two residual blocks for downsampling, while
each decoder layer has three residual blocks for upsampling.
The bottleneck layer mediates between the encoder
and the decoder.
Table \ref{tab:attention-unet} presents the configuration details of our U-Net$^{+}$.
The input and output dimensions of each encoder or decoder layer
are represented as
$H \times W \times C$
where $H$ denotes the height and $W$ denotes the width of the input of an
encoder/decoder layer; $C$ denotes the number of channels the layer has.

It should be pointed out that 
our U-Net$^{+}$ differs from U-Net \citep{falk2019u} in three ways. First, the input of U-Net is an RGB image. 
In contrast, the input of U-Net$^{+}$ contains two magnetogram images, namely the conditional image
$(\mathbf{y}-\mathbf{w})$ and
$\hat{\mathbf{x}}_{t}$ (see Figure \ref{fig:diff_architecture}).
U-Net uses three input channels to read the RGB image, while
U-Net$^{+}$ has two input channels. 
Second, we add self-attention blocks to
U-Net$^{+}$ to help focus on relevant features in the input. However, these self-attention blocks are absent in U-Net. Third, we employ six encoder layers and six decoder layers in U-Net$^{+}$ to improve the model inference and generalization capabilities, 
while U-Net has three encoder layers and three decoder layers. 
During training,
we use a batch size of 64 and a total of 500 epochs
with AdamW being the optimizer, which is an improved version of Adam that decouples weight decay from the gradient update process, leading to better generalization and reduced overfitting \citep{Goodfellow-et-al-2016}.
Table \ref{tab:genmdi-hyperparams} summarizes the hyperparameters used for model training. 
These hyperparameter values are obtained using the grid search capability of the Python machine learning
library, scikit-learn \citep{10.5555/1953048.2078195}.

\begin{table}[]
\begin{tabular}{cccccc}
\hline
Layer      & Number of      & Number of       & Upsampling/  & Input                       & Output                      \\
           & Self-attention Blocks & Residual Blocks & Downsampling & Dimension   
           & Dimension 
           \\ \hline
E1         & 0              & 2               & Down         & $128 \times 128 \times 2$   & $64 \times 64 \times 192$   \\
E2         & 0              & 2               & Down         & $64 \times 64 \times 192$   & $32 \times 32 \times 192$   \\
E3         & 0              & 2               & Down         & $32 \times 32 \times 192$   & $16 \times 16 \times 384$   \\
E4         & 0              & 2               & Down         & $16 \times 16 \times 384$   & $8 \times 8 \times 384$     \\
E5         & 2              & 2               & Down         & $8 \times 8 \times 384$     & $4 \times 4 \times 768$     \\
E6         & 2              & 2               & N/A           & $4 \times 4 \times 768$     & $4 \times 4 \times 768$     \\ \hline
Bottleneck & 1              & 2               & N/A           & $4 \times 4 \times 768$     & $4 \times 4 \times 768$     \\ \hline
D1         & 3              & 3               & Up           & $4 \times 4 \times 768$     & $8 \times 8 \times 768$     \\
D2         & 3              & 3               & Up           & $8 \times 8 \times 768$     & $16 \times 16 \times 768$   \\
D3         & 0              & 3               & Up           & $16 \times 16 \times 768$   & $32 \times 32 \times 384$   \\
D4         & 0              & 3               & Up           & $32 \times 32 \times 384$   & $64 \times 64 \times 384$   \\
D5         & 0              & 3               & Up           & $64 \times 64 \times 384$   & $128 \times 128 \times 192$ \\
D6         & 0              & 3               & N/A           & $128 \times 128 \times 192$ & $128 \times 128 \times 1$   \\ \hline
\end{tabular}
\caption{Configuration Details of 
the U-Net$^{+}$ Used in Our GenMDI Model}
\label{tab:attention-unet}
\end{table}

\begin{table}[]
\centering
\begin{tabular}{lc}
\hline
Parameter & Value \\
\hline
Optimizer & AdamW \\
Weight Decay & $1 \times 10^{-4}$ \\
Learning Rate & $1 \times 10^{-4}$ \\
Batch Size & 64 \\
Epoch & 500 \\
Dropout Rate & 0.1 \\
\hline
\end{tabular}
\caption{Hyperparameters Used for Model Training}
\label{tab:genmdi-hyperparams}
\end{table}

\section{Experiments and Results}
\label{sec:results}
\subsection{Evaluation Metrics}
\label{subsec:metrics}

To evaluate our proposed GenMDI model and compare it with the closely related linear interpolation method, we employed three metrics: 
Pearson's 
correlation coefficient 
\citep[PCC;][]{2024ApJS..271....9D}, 
the peak signal-to-noise ratio
\citep[PSNR;][]{2024SoPh..299...36X}, and changes of magnetic fields \citep[CMF;][]{2019LRSP...16....3T}.
PCC measures the linear correlation between two magnetogram images
$\mathbf{A}$ and $\mathbf{B}$ of equal size, each containing $n$ pixels. It is defined as:
\begin{equation}
\mathrm{PCC}=\frac{\sum_{i=1}^n\left(\mathbf{A}_i-\mu_\mathbf{A}\right)\left(\mathbf{B}_i-\mu_\mathbf{B}\right)}{\sqrt{\sum_{i=1}^n\left(\mathbf{A}_i-\mu_\mathbf{A}\right)^2} \sqrt{\sum_{i=1}^n\left(\mathbf{B}_i-\mu_\mathbf{B}\right)^2}},
\end{equation}
where $\mathbf{A}_i$ and $\mathbf{B}_i$ represent the pixel values at the $i$th position in $\mathbf{A}$ and $\mathbf{B}$, respectively,
and $\mu_\mathbf{A}$ and $\mu_\mathbf{B}$ denote the mean pixel values of $\mathbf{A}$ and $\mathbf{B}$, respectively.
PCC values range from $-1$ to 1,
with $-1$ indicating a perfect negative correlation, 
1 representing a perfect positive correlation, and 0 meaning that there is no correlation.

PSNR, a metric frequently used in 
image super-resolution tasks, is defined as:
\begin{equation}
\operatorname{PSNR}=10 \log _{10}\left(\frac{\mathrm{MAX}^2}{\mathrm{MSE}}\right),
\end{equation}
where MAX represents the maximum possible pixel value fluctuation in a magnetogram and MSE denotes the mean squared error between $\mathbf{A}$ and $\mathbf{B}$. 
We restrict the maximum magnetic field strength of a pixel to 2000 G, and hence MAX is set to 4000 to accommodate both positive and negative magnetic flux regions. 
Higher PSNR values indicate a closer resemblance between $\mathbf{A}$ and $\mathbf{B}$.

The CMF from magnetogram image $\mathbf{A}$ to magnetogram image $\mathbf{B}$ is defined as 
\begin{equation}
\mathrm{CMF} = \frac{1}{n} \sum_{i=1}^n |\mathbf{B}_i - \mathbf{A}_i|.   
\end{equation}
CMF helps to understand the
field variations between two magnetograms.

\subsection{Comparative Analysis}

\begin{figure}
\centering
\includegraphics[width=0.95\linewidth]{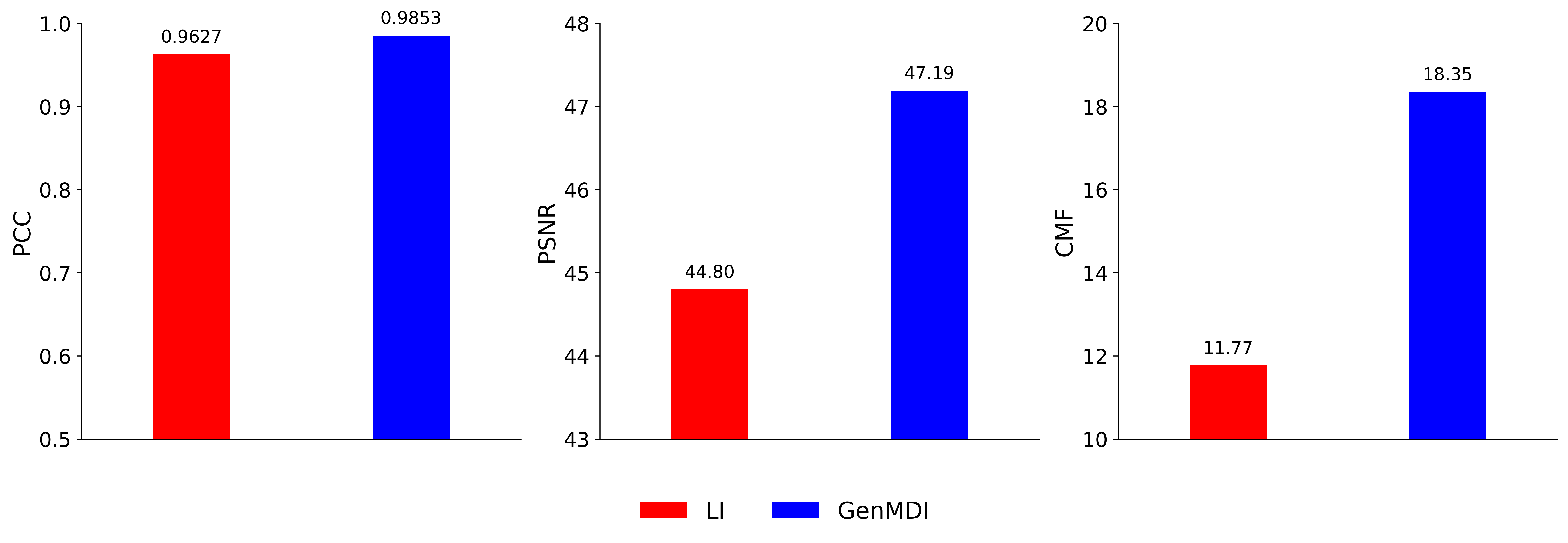} 
\caption{Performance comparison between our GenMDI model and
the linear interpolation (LI) method on the whole test set.
GemMDI performs better than LI in all metrics.}
\label{fig:perf_avg}
\end{figure}

Our GenMDI model is built on top of the
linear interpolation (LI) method.
LI is commonly used to estimate unknown values in a sequence by assuming a linear progression between known data points. In the context of
magnetogram image interpolation, the LI method
constructs new images by linearly blending the
magnetic field strengths of neighboring magnetograms. 
In our study,
for a triplet of images
$(\mathbf{w}, \mathbf{x}, \mathbf{y})$ in the test set,
we first calculate the image
$\mathbf{x}_{I}$ obtained by applying the LI method to
$\mathbf{w}$ and $\mathbf{y}$.
The GenMDI model then predicts a synthetic difference image
$\hat{\mathbf{\mathbf{x}}}_{0}$,
which approximates
the true difference image
$\mathbf{\mathbf{x}}_0$ = 
$(\mathbf{x} - \mathbf{x}_{I})$.
Finally, the model predicts/generates as output
the synthetic image 
$\hat{\mathbf{x}}$
 = $\mathbf{x}_{I}$ + $\hat{\mathbf{\mathbf{x}}}_0$, 
 which approximates the
ground truth image $\mathbf{x}$.

Figure \ref{fig:perf_avg}
compares the mean metric values between
GenMDI and the LI method
on all triplets of images $(\mathbf{w}, \mathbf{x}, \mathbf{y})$ in the test set.
When calculating PCC and PSNR,
we compare the GenMDI-predicted image
(LI-interpolated image, respectively)
with the corresponding true magnetogram image.
When calculating CMF, we compare
$\mathbf{w}$ with $\mathbf{x}$ ($\mathbf{w}$ with the GenMDI-predicted image,
$\mathbf{w}$ with the LI-interpolated image, respectively).
It can be seen in Figure \ref{fig:perf_avg} 
that GenMDI outperforms the LI method
in all metrics. 
Higher PCC scores reflect more accurate predictions of the pixel values relative to the true data.
Higher PSNR values indicate that GenMDI-predicted images have higher fidelity with less distortion than LI-interpolated images.
Overall,
GenMDI improves the LI method
by 2.35\% in the mean PCC and
by 5.33\% in the mean PSNR.
The mean CMF from $\mathbf{w}$ to $\mathbf{x}$ 
($\mathbf{w}$ to the GenMDI-predicted image,
$\mathbf{w}$ to the LI-interpolated image, respectively)
is 23.99 G (18.35 G, 11.77 G, respectively).
On average, GenMDI's CMF values are closer to
the true field variances between magnetograms
than LI's CMF values.
Thus, GenMDI better captures, on average, 
changes in the magnetic field than the LI method.
These findings underscore the capability of our proposed GenMDI model to handle the complex dynamics and intricate details inherent in MDI magnetograms, making it a better choice for
generating synthetic MDI magnetogram images than the traditional linear interpolation method.

\subsection{Case Studies}

\begin{figure}
\centering 
\includegraphics[width=0.99\linewidth]{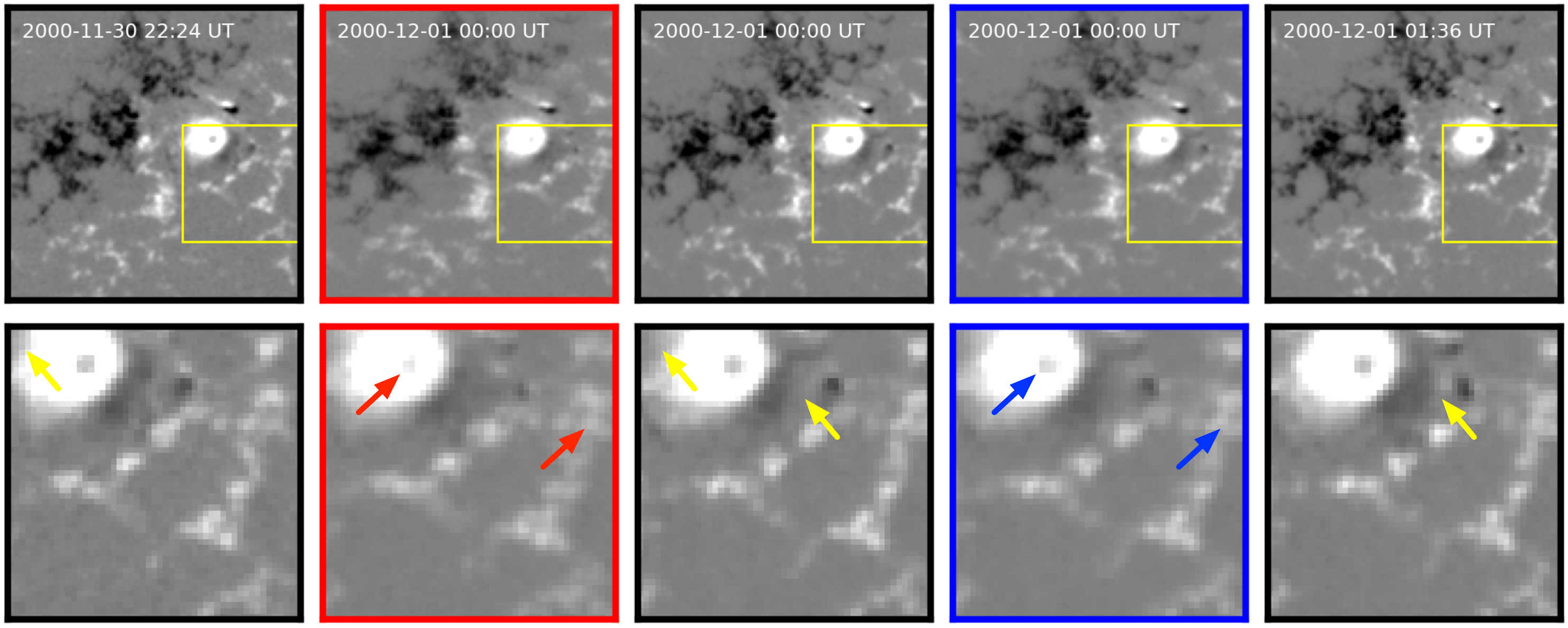}
\caption{Comparison between the magnetograms
produced by the GenMDI model and the LI method on AR 9240. 
(Top panels) 
Three consecutive observed MDI magnetograms, enclosed by black boundary lines,
are taken at 22:24 UT on 30 November 2000, and at 00:00 UT and 01:36 UT on 1 December 2000 respectively, with a cadence of 96 minutes. 
The synthetic magnetogram at 00:00~UT on 1 December 2000, 
enclosed by a red (blue, respectively) boundary line, 
is interpolated by the LI method
(predicted by the GenMDI model, respectively).
(Bottom panels) 
The FOV of the region
highlighted by the yellow box in each corresponding
magnetogram in the top row is displayed.
The areas pointed to by the red and blue arrows  
show differences 
between the 
LI-interpolated 
magnetogram and
GenMDI-predicted magnetogram.
The areas pointed to by the yellow arrows 
highlight differences among the observed magnetograms at different times. 
}
\label{fig:diff_cases1}
\end{figure}

In this section, we present two case studies in which we illustrate
the synthetic magnetograms predicted by our GenMDI model and compare them with the magnetograms interpolated by the LI method.
These two case studies focus on ARs with larger CMF values that have more
changes in the magnetic field.
Figure~\ref{fig:diff_cases1} 
presents results of the first case study on NOAA AR 9240. 
The top row in the figure shows three consecutive observed MDI magnetograms, enclosed by black boundary lines,
collected at 22:24~UT on 30 November 2000, and at 00:00~UT and 01:36~UT on 
1 December 2000 respectively, with a 96-minute cadence. 
The synthetic magnetogram at 00:00~UT 
on 1 December 2000, 
enclosed by a red (blue, respectively) boundary line, 
is interpolated by the LI method
(predicted by the GenMDI model, respectively). 
In comparison of the GenMDI-predicted magnetogram (LI-interpolated magnetogram, respectively) 
with the true magnetogram at 00:00~UT on 1 December 2000, the GenMDI model (LI method, respectively) 
achieves a PCC of 0.9825 (0.9319, respectively) and a PSNR of 40.63 (33.35, respectively). 
Thus, the GenMDI model improves the LI method by 5.43\% in PCC 
and 21.83\% in PSNR.
Moreover, the CMF from the true magnetogram at 22:24~UT 
on 30 November 2000 to the true magnetogram
(GenMDI-predicted magnetogram, LI-interpolated magnetogram, respectively) at 00:00~UT on 1 December 2000
is 88.15~G (68.25 G, 43.77 G, respectively).
In terms of the metric values, the
GenMDI-predicted magnetogram is closer
to the true magnetogram than the LI-interpolated magnetogram.
The bottom row of Figure~\ref{fig:diff_cases1} 
shows the field of view (FOV) of the region
highlighted by the yellow box in each corresponding
magnetogram in the top row.
Comparing the areas pointed to by the arrows in the FOVs,
we see that
the GenMDI-predicted magnetogram is visually also closer
to the true magnetogram than the LI-interpolated magnetogram.

\begin{figure}
\centering 
\includegraphics[width=0.99\linewidth]{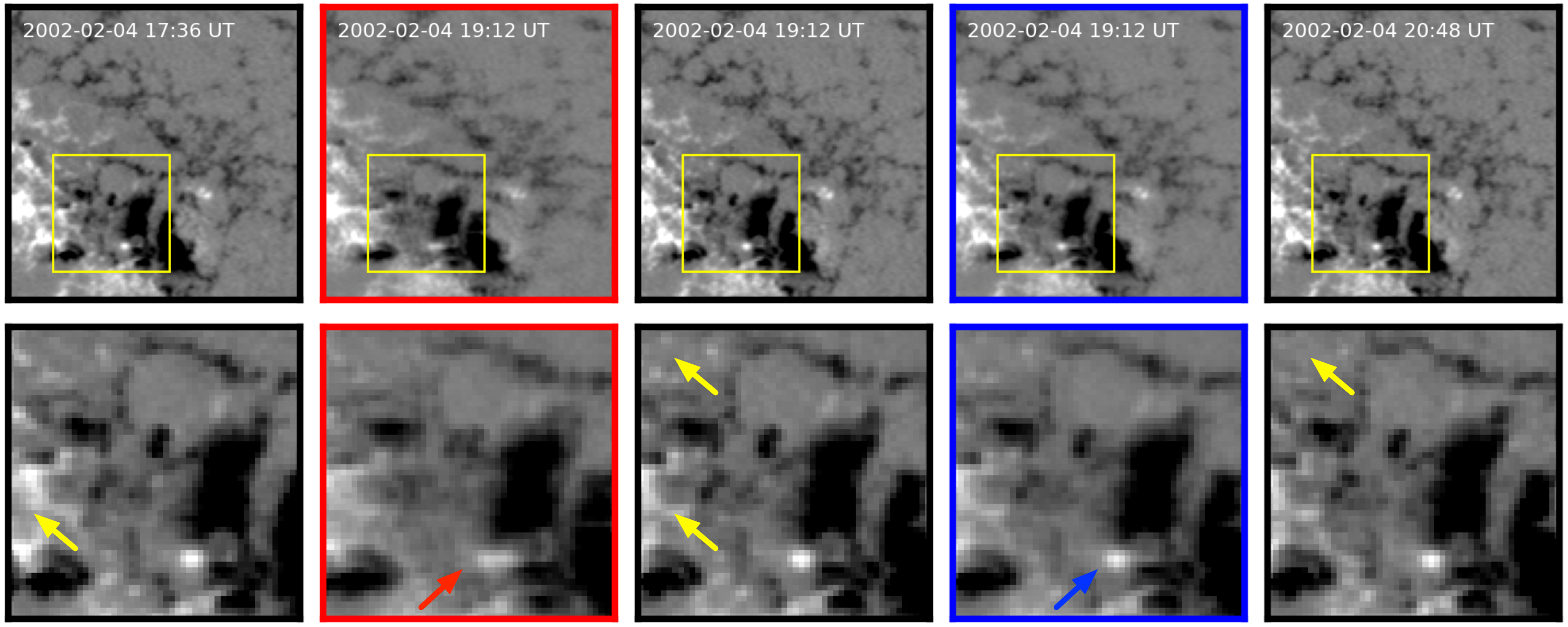}
\caption{Comparison between the magnetograms 
produced by the GenMDI model and the LI method on 
AR 9802. 
(Top panels) Three consecutive observed MDI magnetograms, 
enclosed by black boundary lines, are taken at 17:36 UT, 19:12 UT and 20:48 UT respectively on 4 February 2002, with a cadence of 96 minutes. 
The synthetic magnetogram 
at 19:12 UT on 4 February 2002, 
enclosed by a red (blue, respectively) boundary line, is
interpolated by the LI method
(predicted by the GenMDI model, respectively).
(Bottom panels)
The FOV of the region highlighted by the yellow box in each
corresponding magnetogram in the top row is displayed.
The areas pointed to by the 
red and blue arrows
show differences 
between the 
LI-interpolated 
magnetogram and
GenMDI-predicted magnetogram.
The areas pointed to by the yellow arrows 
highlight differences 
among the observed magnetograms at different times.}
\label{fig:diff_cases2}
\end{figure}

\begin{figure}
\centering 
\includegraphics[width=0.95\linewidth]{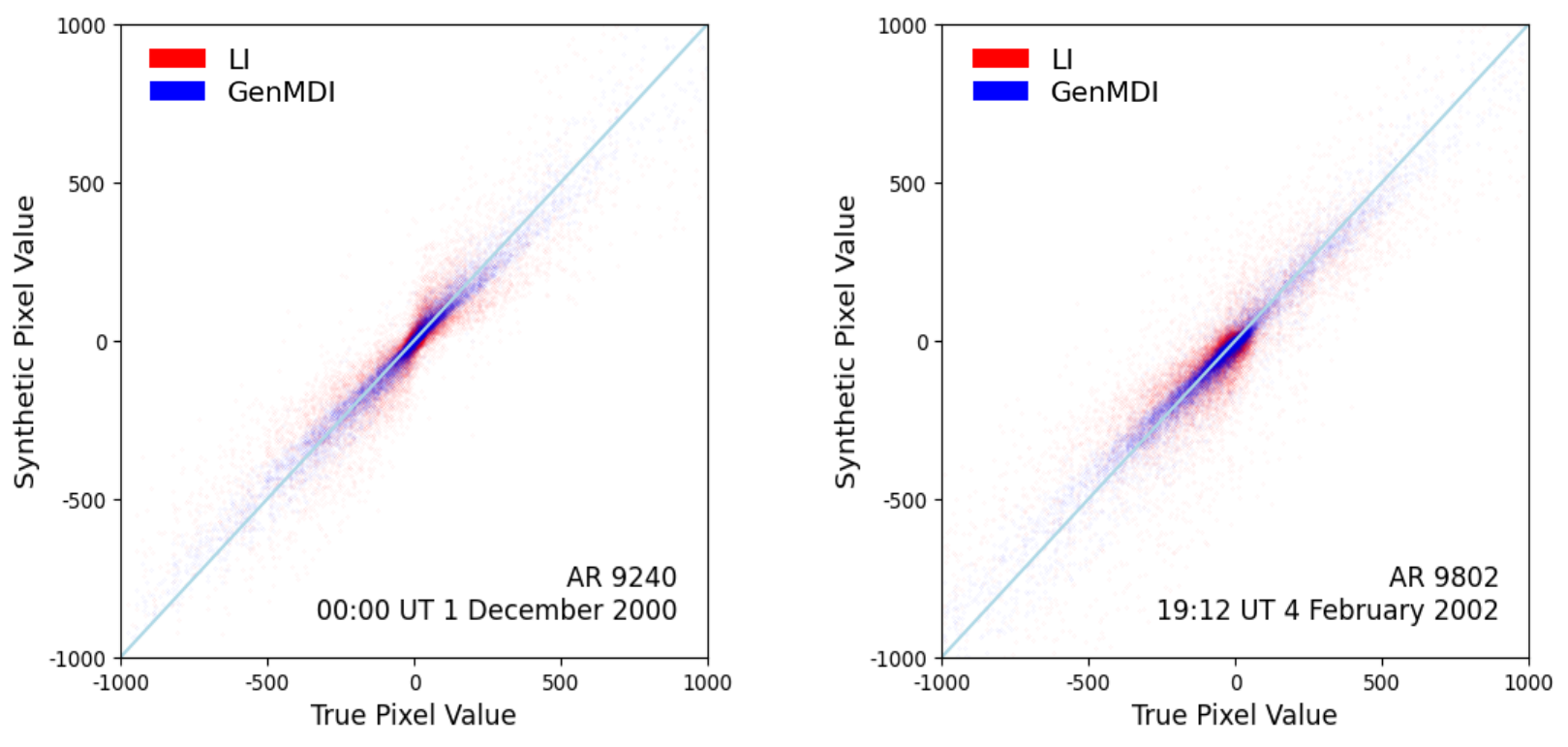} 
\caption{Scatter plots for two ARs. (Left panel) Scatter plot in which the X-axis represents the true pixel values of the observed magnetogram 
and the Y-axis represents the synthetic pixel values of the LI-interpolated (GenMDI-predicted, respectively) magnetogram at 00:00 UT on 1 December 2000 in AR 9240 shown in Figure \ref{fig:diff_cases1}. (Right panel) Scatter plot in which the X-axis represents the true pixel values of the observed magnetogram and the Y-axis represents the synthetic pixel values of 
the LI-interpolated (GenMDI-predicted, respectively) magnetogram at 19:12 UT on 4 February 2002 in AR 9802 shown in Figure \ref{fig:diff_cases2}. 
In both ARs, GenMDI-predicted magnetograms are closer to the observed/true MDI magnetograms than the LI-interpolated magnetograms.}
\label{fig:scatter_mdi}
\end{figure}

Figure~\ref{fig:diff_cases2} 
presents results of the second case study on NOAA AR 9802. 
The top row in the figure
shows three consecutive observed MDI magnetograms, 
enclosed by black boundary lines, collected at 
17:36~UT, 19:12~UT, and 20:48~UT on 4 February 2002 respectively. 
The synthetic magnetogram at 
19:12~UT, enclosed by a red (blue, respectively)
boundary line, is 
interpolated by the LI method
(predicted by the GenMDI model, respectively). 
In comparison of the GenMDI-predicted magnetogram
(LI-interpolated magnetogram) with the true magnetogram at 19:12 UT, 
the GenMDI model (LI method, respectively) achieves 
a PCC of 0.9797 (0.9103, respectively) and a 
PSNR of 39.02 (31.41, respectively). 
Thus, the GenMDI model improves the LI method by 
7.62\% in PCC 
and 24.23\% in PSNR. 
Furthermore, the CMF from the true magnetogram at 17:36 UT to the 
true magnetogram (GenMDI-predicted magnetogram, LI-interpolated magnetogram, respectively) 
at 19:12~UT is 120.88 G (88.47 G, 61.02 G).
The bottom row of Figure~\ref{fig:diff_cases2} 
shows the FOV of the region highlighted by the yellow box in each
corresponding magnetogram in the top row.
It can be seen that the GenMDI-predicted 
magnetogram is closer to the true magnetogram
than the LI-interpolated magnetogram. 
Figure \ref{fig:scatter_mdi} presents scatter plots,
which further support this conclusion.
It is worth pointing out that the PCC and PSNR improvement rates in the two case studies are higher than those in
Figure \ref{fig:perf_avg}, 
indicating that our GenMDI model performs much better than the LI method
in ARs with larger CMF values having more changes in the magnetic field.

\subsection{Temporal Super-resolution of 
MDI Magnetograms}

So far, we have shown how to use
our GenMDI tool to create a synthetic magnetogram
between two observed magnetograms.
We can extend this approach
by iteratively applying the tool to
two neighboring magnetograms to enhance the temporal resolution of MDI
magnetograms from 96 to 12 min.
Specifically, consider two adjacent
images $\mathbf{w}$ and $\mathbf{y}$
in a sequence of MDI
magnetogram images.
We generate seven synthetic images
between $\mathbf{w}$ and $\mathbf{y}$ as follows.
We first generate a synthetic image
$\mathbf{x}_{4}$ between $\mathbf{w}$ and $\mathbf{y}$ using the trained GenMDI model.
Then, we use the model to generate a synthetic image
$\mathbf{x}_{2}$ between $\mathbf{w}$ and $\mathbf{x}_{4}$.
Next, we use the model to generate a synthetic image
$\mathbf{x}_{1}$ between $\mathbf{w}$ and $\mathbf{x}_{2}$.
Finally, we use the model to generate a synthetic image
$\mathbf{x}_{3}$ between $\mathbf{x}_{2}$ and $\mathbf{x}_{4}$.
This process creates three synthetic images
$\mathbf{x}_{1}, \mathbf{x}_{2}, \mathbf{x}_{3}$ between $\mathbf{w}$ and $\mathbf{x}_{4}$.
Similarly, we can create three synthetic images
$\mathbf{x}_{5}, \mathbf{x}_{6}, \mathbf{x}_{7}$ between
$\mathbf{x}_{4}$ and $\mathbf{y}$.
In total, we generate seven synthetic images
$\mathbf{x}_{1}, \ldots , \mathbf{x}_{7}$
between $\mathbf{w}$ and $\mathbf{y}$ to obtain a new sequence of images.
The original sequence of MDI
magnetogram images has a cadence of 96 minutes.
With the synthetic images generated, 
the new sequence of images has a cadence of 12 minutes.
The time stamp of $\mathbf{x}_{i}$, $1 \leq i \leq 7$, is equal to the time stamp of $\mathbf{w}$ plus $i \times 12$ minutes.

\begin{figure}
\centering 
\includegraphics[width=0.99\linewidth]{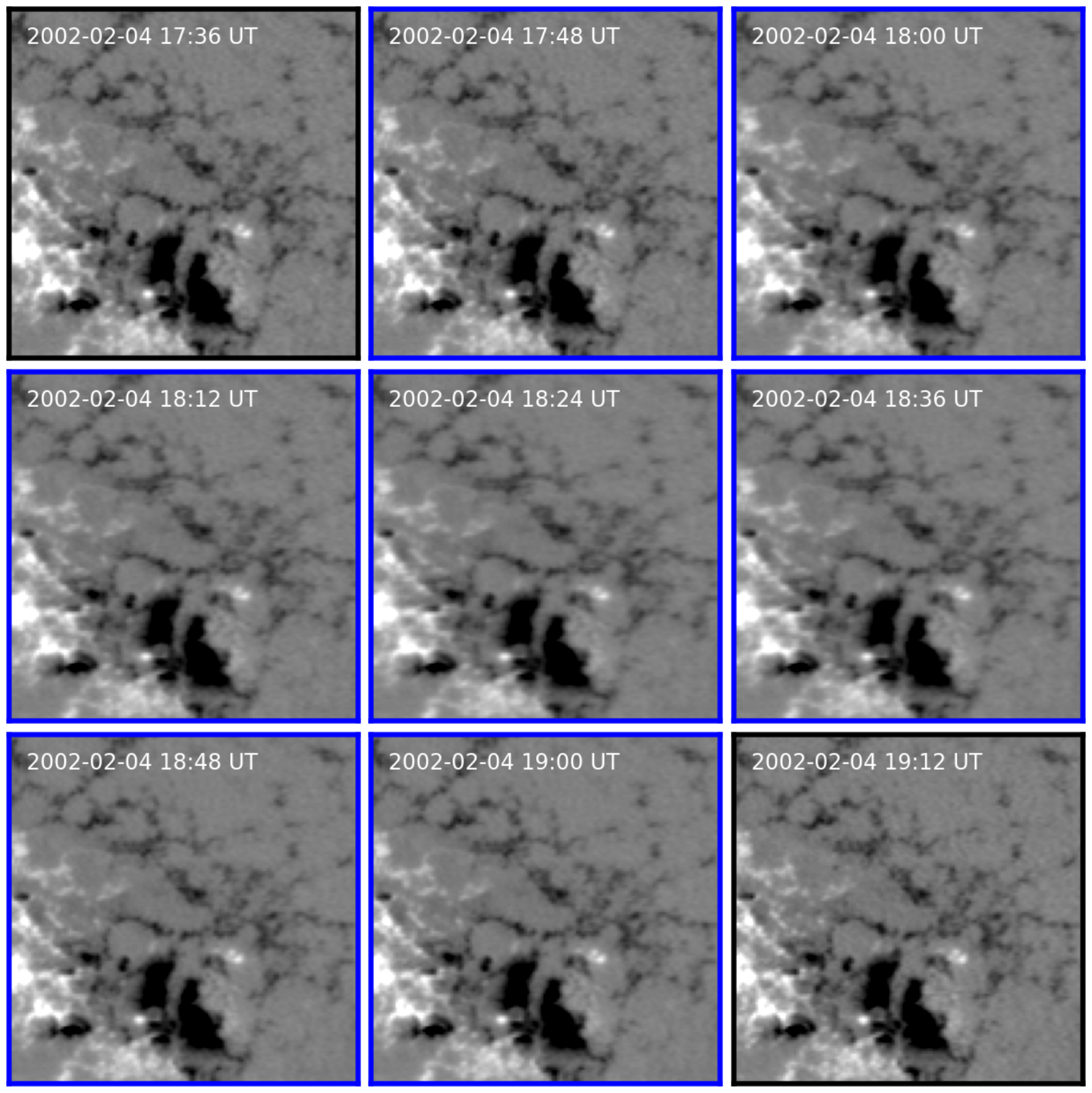}
\caption{Illustration of temporally super-resolved MDI magnetograms.
Observed magnetograms are enclosed by black boundary lines,
while synthetic magnetograms predicted by GenMDI are enclosed by
blue boundary lines.
The observed magnetograms
are taken from
AR 9802, collected at 17:36 UT and 19:12 UT, respectively,
on 4 February 2002.
 The observed magnetograms, 
 together with the seven synthetic magnetograms,
 occur at a cadence of 12 min.}
\label{fig:superresolution}
\end{figure}

Figure \ref{fig:superresolution} shows
the temporal super-resolution results in AR 9802
where the first and the last 
observed magnetograms, enclosed by 
black boundary lines, 
together with the seven GenMDI-predicted 
synthetic magnetograms,
enclosed by blue boundary lines,
occur at a cadence of 12 min.
The two observed magnetograms
are collected at 17:36 UT and 19:12 UT,
respectively, on 4 February 2002.
The synthetic magnetograms
help scientists better understand the
dynamics and evolution of ARs
on a finer scale.
With the proposed GenMDI tool, we can
create a temporally consistent set of LOS magnetograms
between SDO/HMI ({\sf hmi.sharp\_cea\_720s} data series at JSOC) and 
SOHO/MDI ({\sf mdi.smarp\_cea\_96m} data series at JSOC),
at the same cadence of 12 min.

To validate our approach,
we collect ARs in the overlap period between 
MDI ({\sf mdi.smarp\_cea\_96m} data series) and HMI
({\sf hmi.sharp\_cea\_720s} data series).
In this overlap period, namely year 2010 in the two data
series, we select 15 ARs, which
contain temporally and spatially matched LOS
magnetograms between MDI and HMI.
In each AR, we select two neighboring MDI magnetograms,
$\mathbf{x}_{\mbox{\tiny MDI}}^{1}$ and
$\mathbf{x}_{\mbox{\tiny MDI}}^{2}$,
and the corresponding HMI magnetograms
$\mathbf{x}_{\mbox{\tiny HMI}}^{1}$ and
$\mathbf{x}_{\mbox{\tiny HMI}}^{2}$ that
temporally and spatially match
$\mathbf{x}_{\mbox{\tiny MDI}}^{1}$ and $\mathbf{x}_{\mbox{\tiny MDI}}^{2}$,
respectively.
We reduce the spatial resolution of the HMI magnetograms
by a factor of 4 so that the HMI and MDI magnetograms
have the same spatial resolution.
We then use the GenMDI model
(the LI method, respectively)
to create seven synthetic MDI magnetograms
between $\mathbf{x}_{\mbox{\tiny MDI}}^{1}$ and
$\mathbf{x}_{\mbox{\tiny MDI}}^{2}$.
The mean PCC (PSNR, respectively)
between the true/observed magnetograms
$\mathbf{x}_{\mbox{\tiny MDI}}^{1}$ and $\mathbf{x}_{\mbox{\tiny HMI}}^{1}$
as well as $\mathbf{x}_{\mbox{\tiny MDI}}^{2}$ and $\mathbf{x}_{\mbox{\tiny HMI}}^{2}$
in the 15 ARs is
0.8462 (32.11, respectively).
The mean PCC
between the seven GenMDI-predicted
(LI-interpolated, respectively)
MDI magnetograms
and the corresponding true/observed HMI magnetograms
in the 15 ARs is
0.7817 
(0.7626, respectively).
The mean PSNR
between the seven GenMDI-predicted
(LI-interpolated, respectively)
MDI magnetograms
and the corresponding true/observed HMI magnetograms
in the 15 ARs is
30.31
(29.58, respectively).
Overall, 
the GenMDI-predicted MDI magnetograms
are closer to
the corresponding true/observed HMI magnetograms
than the
LI-interpolated MDI magnetograms.

\section{Conclusion}
\label{sec:conclusion}

We present GenMDI, a deep generative model
for improving the temporal resolution of SOHO/MDI magnetograms.
Using a conditional diffusion process, GenMDI can create
synthetic magnetograms between observed magnetograms,
allowing scientists to better understand
the changes of magnetic fields, as well as
the dynamics and evolution of
active regions on a finer scale.
Compared to the commonly used linear interpolation method,
GenMDI shows better performance, 
especially in
ARs with
dynamic evolution (large changes) in magnetic fields.
 
Our GenMDI model was trained by
MDI magnetograms with a cadence of 96 min.
In additional experiments,
we trained the model using
MDI magnetograms with a cadence of 192
or 384 min.
The results obtained were consistent with
those reported here.
Thus, we conclude that GenMDI is a feasible
tool for improving the temporal resolution of MDI LOS magnetograms
of solar ARs.
Combining GenMDI with the tools to improve
the spatial resolution
of MDI LOS magnetograms
\citep{2024ApJS..271....9D,
2024ApJS..271...46M,2024SoPh..299...36X}
and to generate MDI vector magnetograms
\citep{2023SoPh..298...87J},
we are able to create a uniform set of
data, close to HMI quality, covering
the period from 1996 to the present.
This uniform data set would provide valuable insight into
the study of dynamics and evolution of solar magnetic fields in ARs.

In a recent study, \citet{2024A&A...686A.272S} combined diffusion models with generative adversarial networks (GANs) to improve the spatial resolution of HMI continuum images. Our previous work \citep{2023SoPh..298...87J} employed a convolutional neural network to generate MDI vector magnetograms. In the future, it will be worthwhile to investigate the use of diffusion models combined with GANs to generate MDI vector magnetograms and to improve the spatial and temporal resolutions of the generated MDI vector data.\\

\noindent
The authors thank members of the Institute for Space Weather Sciences
for fruitful discussions.
V.Y. acknowledges support from NSF grants
AST-2108235,
AGS-2114201,
AGS-2300341, and
AGS-2309939.
J.W. and H.W. acknowledge support from NSF grants
AGS-2149748, AGS-2228996, OAC-2320147, and NASA grants 80NSSC24K0548,
80NSSC24K0843, and 80NSSC24M0174.
K.A. acknowledges support from King Saud University, Saudi Arabia.
Y.X. acknowledges support from NSF grants
AGS-2228996,
AGS-2229064, and
RISE-2425602.

\bibliographystyle{aasjournal}

\end{document}